\documentclass[adp,fleqn]{w-art}
\usepackage{times}
\usepackage{w-thm}
\usepackage{amssymb}
\theoremstyle{plain}

\theoremstyle{definition}

\usepackage[]{graphicx}
\chardef\bslash=`\\ 

\newcommand{\md}{{\mathrm d}}
\newcommand{\lP}{\ell_{\rm P}}
\newcommand{\sgn}{{\rm sgn}}

\def\rightmark{}

\begin{document}
\DOIsuffix{theDOIsuffix}
\Volume{12}
\Issue{1}
\Copyrightissue{01}
\Month{01}
\Year{2003}
\pagespan{1}{}



\title[Loop quantum cosmology]{Universe scenarios from loop quantum cosmology}


\author[M.\ Bojowald]{Martin Bojowald\footnote{
     e-mail: {\sf mabo@aei.mpg.de}}\inst{1}\inst{2}} 
\address[\inst{1}]{Max-Planck-Institute for Gravitational Physics, Albert-Einstein-Institute, Am M\"uhlenberg 1, 14476 Potsdam, Germany}
\address[\inst{2}]{Institute for Gravitational Physics and Geometry, The Pennsylvania State University, 104 Davey Lab, University Park, PA 16802, USA}
\begin{abstract}
  Loop quantum cosmology is an application of recent developments for
  a non-perturbative and background independent quantization of
  gravity to a cosmological setting. Characteristic properties of the
  quantization such as discreteness of spatial geometry entail
  physical consequences for the structure of classical singularities
  as well as the evolution of the very early universe. While the
  singularity issue in general requires one to use difference
  equations for a wave function of the universe, phenomenological
  scenarios for the evolution are based on effective equations
  implementing the main quantum modifications. These equations show
  generic bounces as well as inflation in diverse models, which have
  been combined to more complicated scenarios.
\end{abstract}
\maketitle                   





\section{Introduction}

The universe is, on large scales, well described by general relativity
which provides the basis for mathematical models of the possible
behavior of a universe. From solutions of Einstein's field equations
one obtains the geometry of space-time once the matter content has
been specified. On this classical level, however, the description will
always remain incomplete as a consequence of singularity theorems: any
space-time evolved backward in time from the conditions we perceive
now will reach a boundary in a finite amount of proper time. At this
boundary, the equations of the theory, and thus classical physics,
break down, often accompanied by curvature divergence. For the
development of complete universe scenarios the classical theory of
general relativity thus has to be extended.

Such an extension is often expected to come from quantum gravity,
where not only general relativistic but also quantum effects are taken
into account. One approach, which is background independent and
non-perturbative and can deal with those extreme conditions realized
at classical singularities, is loop quantum gravity
\cite{Rev}.  It is a canonical quantization and turns
the classical metric and extrinsic curvature into operators on a
Hilbert space. The classical geometrical structures are thus replaced
by properties of operators which by itself leads to a different
formulation. While classical geometry has to re-emerge as an
approximation on large scales, on small scales quantum behavior has to
be taken into account fully. (For general aspects of quantum theory in
the context of cosmology see \cite{CK}.) Then also the singularity
problem appears in a different light as the basic object is not a
space-time metric which cannot be extended beyond singularities in the
classical evolution but a wave function. The wave function, also, is
subject to equations which need to be analyzed in order to see if it
always gives a complete solution telling us what happens at and beyond
classical singularities.

Classical space-time is thus replaced by quantum space-time whose
effects are most important on small scales such as those of a small
universe close to a classical singularity. A fully quantized system is
indeed necessary to describe states right at a classical singularity,
but this is in general complicated at technical as well as conceptual
levels. Close to classical singularities it is thus helpful to have
effective systems which are of classical type, removing
interpretational issues of quantum theories, but take into account
some quantum effects. With those ingredients it is then possible to
develop several complete scenarios for universes, and at the same time
obtain potentially observable phenomenological effects.

\section{Loop quantum cosmology}

Just as Wheeler--DeWitt models, loop quantum gravity is a canonical
quantization, i.e.\ it starts with a foliation of space-time into
spatial slices $\Sigma_t$. Canonical variables are the spatial metric
$q_{ab}$ and its momenta related to extrinsic curvature $K_{ab}$ of
the slice \cite{ADM}. The canonical Poisson algebra then is to be
turned into a suitable operator algebra and represented on a Hilbert
space. As always in field theories, however, operators for the field
values in single points do not exist and the fields have to be smeared
first by integrating them over extended regions. For field theories
other than gravity, this is usually done on 3-dimensional regions
using the background metric to define an integration measure. But for
gravity, the metric itself is dynamical and thus to be smeared, and
using an additional metric for smearing would introduce a background.
A different procedure is required for gravity, i.e.\ we need to smear
fields but do so in a background independent manner.

\subsection{Holonomy-flux algebra}

Indeed, the success of a quantization procedure often depends on the
choice of basic variables. For gravity, it is most helpful to
transform to new basic fields, given by {\em Ashtekar variables}
\cite{AshVar,AshVarReell}.  These variables are also canonical, but
instead of the spatial metric one uses the {\em densitized triad}
$E^a_i$, related to the spatial metric by $E^a_iE^b_i=\det q q^{ab}$,
and the {\em Ashtekar connection} $A_a^i=\Gamma_a^i+\gamma K_a^i$
defined with the spin connection $\Gamma_a^i$ and extrinsic curvature
components $K_a^i$. In addition, there is the Barbero--Immirzi
parameter $\gamma>0$ \cite{AshVarReell,Immirzi}, which for simplicity
will be set equal to one in what follows. (Its value can be computed
from black hole entropy and is smaller than but of the order of one
\cite{LoopEntro}.) In those variables, the
extrinsic curvature term in $A_a^i$ makes it canonically conjugate to
the triad, and $\Gamma_a^i$ provides the transformation properties of
a connection. For $E^a_i$, the important properties compared to the
metric are that it also knows about the orientation of space since it
can be left- or right-handed, and that it is dual to a 2-form
$\epsilon_{abc}E^c_i$. The orientation will be essential later on in
the discussion of singularities, while the transformation properties
of a connection and dual 2-form, respectively, are important right now
because they allow a natural smearing without introducing a background
metric.

We can then integrate a connection along curves, where for good gauge
properties we also take the path ordered exponential, i.e.\ use
{\em holonomies} 
\begin{equation}
 h_e(A)={\cal P}\exp\int_e\tau_i A_a^i\dot{e}^a{\rm d}t
\end{equation}
for curves $e\subset\Sigma$ and
{\em fluxes}
\begin{equation}
  F_S(E)=\int_S \tau^i E^a_i\epsilon_{abc}{\rm d}y^a{\rm d}y^b 
\end{equation}
for surfaces $S\subset \Sigma$. (SU(2)-generators
$\tau_j=-\frac{1}{2}i\sigma_j$ with Pauli matrices $\sigma_j$ appear
because this is the gauge group of triad-rotations not changing the
metric.) There are then no 3-dimensional smearings, but a 1- and a
2-dimensional one, which, as it turns out, adds up to the right
overall smearing for a well-defined quantization.

This is the basis of the background independent quantization provided
by loop quantum gravity \cite{LoopRep}, and it has many crucial
properties as direct consequences.  First, the holonomy-flux algebra,
defined by the Poisson relations, is, under weak mathematical
conditions, represented {\em uniquely} on a Hilbert space together
with a unitary action of diffeomorphisms of $\Sigma$
\cite{FluxAlg}. The latter property is
required for the independence of gravity under the choice of spatial
coordinates. This representation is {\em cyclic}, i.e.\ there is a
basic state from which all others can be obtained by repeatedly acting
with operators of the algebra. Other characteristic properties of the
basic holonomy and loop operators are then very different from those
in a Wheleer--DeWitt quantization: There is no operator for connection
components or even their integrals, but {\em only for holonomies}
unlike in a Wheeler--DeWitt quantization where extrinsic curvature
components are basic operators.  Moreover, flux operators have {\em
discrete spectra} and so do geometrical operators such as area and
volume \cite{AreaVol}. With these properties, one can then
use the basic representation to construct classes of well-defined
Hamiltonian constraint \cite{QSDI}and matter Hamiltonian operators
\cite{QSDV}.

\subsection{Isotropic quantum cosmology}

These techniques also make a symmetry reduction possible which is much
closer to the full theory than a Wheeler--DeWitt model would be
\cite{SymmRed}. When symmetries are imposed, natural sub-algebras of
the full holonomy-flux algebra are defined which, using cyclicity of
the representation, {\em induce} the basic representation of a model.
Here, one simply acts only with those operators in the distinguished
sub-algebra for a given symmetry, and thus obtains less states than
one would get from the full holonomy-flux algebra. In this sense, the
basic representation of models, which is so important for other
physical properties, is directly obtained from the full quantum
theory: quantization is done {\em before} performing the symmetry
reduction, at least as far as the basic representation is concerned.
More complicated operators such as the Hamiltonian constraint can then
be constructed from the basic ones following the steps in the full
theory by analogy.

The simplest case, isotropy, serves as a good example to illustrate
the basic properties \cite{IsoCosmo}.  Classically, there is a single
gravitational degree of freedom, the scale factor $a$ with its
momentum $p_a=-3(4\pi G)^{-1}a\dot{a}$ ($G$ being the gravitational
constant). These are components of the spatial metric and extrinsic
curvature, which now are replaced by Ashtekar variables. In the
isotropic case, there is again a single canonical pair $(c,p)$ with
\begin{equation}
 |p|=a^2 \qquad,\qquad {\rm sgn}(p)\mbox{: orientation}
\end{equation}
and
\begin{equation}
 c={\textstyle\frac{1}{2}}(k+\dot{a}) \qquad,\qquad \mbox{$k=0$: flat, $k=1$:
closed.}
\end{equation}

These variables could be quantized directly in a Wheeler--DeWitt
manner, i.e.\ $a$ as a multiplication operator and
$\hat{p}_a=-i\hbar\partial/\partial a$ on square integrable functions
(of only positive $a$, which means that $\hat{p}_a$ in this manner is
not self-adjpoint), but an analog of the latter operator does not
exist in the full theory.  Through the induced holonomy-flux
representation it is rather the exponentials $\exp(i\mu c/2)$ for any
real $\mu$ (to be thought of as related to the parameter length
of a curve) which are basic in addition to the densitized triad
component $p$ proportional to a flux. The induced representation is
then defined on a Hilbert space with orthonormal basis of states
\begin{equation}
 \langle c|\mu\rangle= e^{i\mu c/2}\qquad, \qquad \mu\in{\mathbb R}
\end{equation}
and basic operators
\begin{eqnarray}
 \hat{p}|\mu\rangle &=& {\textstyle\frac{1}{6}}\lP^2\mu|\mu\rangle\\
 \widehat{e^{i\mu'c/2}}|\mu\rangle&=& |\mu+\mu'\rangle\,. \label{expc}
\end{eqnarray}
Note that $\hat{p}$ is self-adjoint and $\widehat{e^{i\mu'c/2}}$ is
unitary since the full range of real numbers for $\mu$ is allowed
taking into account the orientation freedom in a triad.  In this
representation, basic operators of the model have the same properties
as those in the full theory \cite{Bohr}: $\hat{p}$ has normalizable
eigenstates and thus a discrete spectrum (nonetheless, the set of
eigenvalues is the full real line, which is not in conflict with the
discreteness of the spectrum because the Hilbert space is
non-separable) and one can see that there is no operator for $c$ but
only for exponentials $e^{i\mu c/2}$ not being continuous in
$\mu$.

Compared to a Wheeler--DeWitt quantization, however, the properties
are very different. In that case, the scale factor, related to $p$,
would have a continuous spectrum and extrinsic curvature $p_a$,
related to $c$, would be represented directly as an operator. The
properties realized in loop quantum cosmology as in the full theory
are a consequence of strong restrictions coming from background
independence and its transfer to symmetric models.

On this basic representation we can construct more complicated
operators, most importantly the Hamiltonian constraint. Since the
properties of basic operators are as in the full theory, the
construction can be done in an analogous manner, only adapting to the
symmetric context where needed. Properties of composite operators are
then also close to those in the full theory, although for them the
relation, as of now, is not as tight and there is no derivation of
symmetric composite operators from the full ones.

For the Hamiltonian constraint, we start from the classical Friedmann
equation
\[
 H=-6\left[2c(c-k)+k^2\right]\sqrt{|p|}+8\pi GH_{\rm
matter}(p,\phi,p_{\phi})=0
\]
with matter Hamiltonian $H_{\rm matter}$, e.g.\ 
\begin{equation}\label{Hmatter}
 H_{\phi}=\frac{1}{2}a^{-3}p_{\phi}^2+ a^3V(\phi)
\end{equation}
for a scalar $\phi$ with momentum $p_{\phi}$ and potential $V(\phi)$,
written down in isotropic Ashtekar variables. This is, using the
volume operator $\hat{V}=|\hat{p}|^{3/2}$ with eigenvalues
\begin{equation}
 V_{\mu}=(\ell_{\rm P}^2|\mu|/6)^{3/2}
\end{equation}
and replacing factors of $c$ by exponentials (\ref{expc}), quantized
to an operator equation $\hat{H}|\psi\rangle=0$ for states
$|\psi\rangle=\sum_{\mu}\psi_{\mu}(\phi)|\mu\rangle$ given by a {\em
difference equation} \cite{IsoCosmo,Closed,Bohr} such as
\begin{eqnarray} \label{DiffEq}
&&
    (V_{\mu+5}-V_{\mu+3})e^{ik}\psi_{\mu+4}(\phi)- (2+k^2)
(V_{\mu+1}-V_{\mu-1})\psi_{\mu}(\phi)\\\nonumber
&&+
    (V_{\mu-3}-V_{\mu-5})e^{-ik}\psi_{\mu-4}(\phi)
  = -{\textstyle\frac{4}{3}}\pi
G\ell_{\rm P}^2\hat{H}_{\rm matter}(\mu)\psi_{\mu}(\phi)
\end{eqnarray}
for $\psi_{\mu}(\phi)$. Unlike Wheeler--DeWitt equations which are
differential, there is thus a difference equation as a result of
discrete quantum geometry. Nevertheless, for large $\mu\gg1$ the
difference operators can be expanded in a Taylor series provided that
the wave function is sufficiently differentiable
\cite{SemiClass,Bohr}. This property can be included in
semiclassicality conditions \cite{DynIn,FundamentalDisc}, but in more
complicated models the existence of sufficiently differentiable
solutions may not be guaranteed \cite{GenFunc}.

\subsection{Classical singularities}
  
With the propagation equation for the wave function on minisuperspace
we can now address the issue of singularities in quantum
cosmology. Compared to the classical situation, the setup has changed
since we do not have evolution equations for the metric in coordinate
time, but an equation for the wave function on minisuperspace. In
semiclassical regimes, the metric can be reconstructed from the wave
function, for instance making use of observables (as completed for a
free, massless scalar in \cite{APS}). But it is not guaranteed that a
classical geometric picture is suitable everywhere for the behavior of
a universe. From the point of view of quantum gravity, the propagation
equation for the wave function is more fundamental, and so also the
singularity issue is to be addressed at this level. The question then
arises whether or not initial values in one classical regime of large
volume are sufficient to determine the quantum solution on all of
(mini)superspace including regions which can be interpreted as being
beyond a classical singularity. If this is the case, quantum
space-time would not be incomplete and thus non-singular.

For isotropic quantum cosmology, this question can be answered
immediately. Minisuperspace is, first of all, enlarged compared to the
usual metric space in that we have two regions differing by their
orientation $\sgn(p)$ and separated by degenerate geometries.
Classically, they are thus separated by singularites. In quantum
theory, the configuration space for wave functions $\psi_{\mu}(\phi)$
also knowns about orientation via $\sgn(\mu)$, and a classical
singularity would occur at $\mu=0$. Unlike the classical evolution,
the quantum equation for the wave function uniquely yields the wave
function on one side if we pose initial values on the other side. In
this way, aspects of quantum gravity give us, first, a new region of
minisuperspace and, second, a unique extension between the two
regions. In this manner, quantum gravitational models are {\em
  singularity-free} \cite{Sing}. So far, explicit constructions
include anisotropic (Bianchi class A) models \cite{HomCosmo,Spin} and
as inhomogeneous cases spherical symmetry and polarized cylindrical
gravitational waves \cite{SphSymm,SphSymmSing}.

The criterion of extendability used here is the most direct and most
general one for the singularity issue, both at the classical as well
as quantum level. In quantum cosmology, other criteria have been
used such as the finiteness of the wave function at a classical
singularity. Also this is realized automatically in loop quantum
cosmology because, thanks to discreteness, there is always only a
finite number of computational steps between initial values and a
classical singularity. Infinities in the wave function could then only
arise from diverging matter Hamiltonians, which does not occur in loop
quantizations as we will explain more explicitly in what follows. In
this context, we will also discuss curvature divergence which
sometimes serves as a singularity criterion.

At a more intuitive level, we obtain the picture of a universe which
starts in a collapsing branch and evolves through a phase where
continuous geometry fails but discrete quantum geometry remains
meaningful. At the transition, the orientation of space changes
implying that the universe turns its inside out. What exactly happens
during the transition depends on the concrete matter model. If one has
a parity violating matter Hamiltonian such as that of the standard
model, there are changes between the two branches even in the
equations of motion.  Otherwise, the two sides generically are still
different from each other depending on the initial conditions for the
wave function. Even if one starts close to a classical geometry on one
side, it may then happen that after evolving through a violent quantum
regime a new classical geometry will not be recovered. This does,
however, not happen in matter models with a free, massless scalar
which can be treated completely \cite{APS}. In this case, solutions
which are semiclassical for large volume bounce at small volume and
become semiclassical afterwards. For those solutions, there is no
difference in behavior when the orientation is changed.

As already noted, the difference equation can be well approximated by
the usual Wheeler--DeWitt equation on large scales.  On small scales,
the equations differ considerably, but in some cases one can still
compare the effects of initial conditions. For the Wheeler--DeWitt
equation, initial conditions have originally been imposed at $a=0$
with the intention of removing the classical singularity by requiring
the wave function to vanish there
\cite{DeWitt,SIC}. However, this turns out to be ill-posed as an
initial value problem, i.e.\ in most models only the trivial solution
of a vanishing wave function exists. In loop quantum cosmology, on the
other hand, {\em dynamical initial conditions} have been derived from
the difference equation \cite{DynIn,Essay} which in many cases are
comparable to DeWitt's condition $\psi(0)=0$ but making it well-posed
in the discrete setting \cite{Scalar}. For a closed model, one can see
\cite{BoundProp} that the dynamical initial conditions are closer to
the no-boundary proposal \cite{nobound} than to the tunneling proposal
\cite{tunneling}.  But also in the discrete setting, the situation is
more complicated in less symmetric models such as anisotropic
\cite{GenFunc} or inhomogeneous ones
\cite{SphSymmSing} where analogous mechanisms are more difficult to
realize. The possibility of conditions following from the constraint
equation also depends on the ordering of the Hamiltonian constraint
operator which may change coefficients in the difference equation. The
ordering for (\ref{DiffEq}) as also used in \cite{DynIn} is not
symmetric, and using a symmetric one removes additional conditions for
wave functions in isotropic models. Symmetric orderings are often
required for semiclassical issues or for computing the physical inner
product, and moreover for a non-singular evolution in inhomogeneous
models \cite{SphSymmSing}. In such a situation, isotropic models loose
their dynamical initial conditions, but some conditions remain in
anisotropic models \cite{BHInt} as well as the inhomogeneous
case. Here, however, the analysis of implications for the solution
space is still very incomplete.  A general mechanism to provide
initial conditions for the wave function of a universe is thus still
outstanding.

\subsection{Matter Hamiltonian}

For a complete cosmological model, we also need to know its matter
content and quantize its Hamiltonian for quantum cosmology. This, now,
does not only include the matter fields but also geometrical factors
in a matter Hamiltonian which need to be turned into operarors. For
instance for a scalar, the Hamiltonian
$H_{\phi}=\frac{1}{2}{a^{-3}}p_{\phi}^2+a^3 V(\phi)$
has to become an operator in the field values $\phi$ but also in the
scale factor $a$. We thus need a quantization of the inverse of $a$,
or $p$, in order to quantize the kinetic term.

At this point, properties of the basic representation become
important: As we have seen, a loop quantization leads to an operator
$\hat{p}$ which has a discrete spectrum containing zero, and such an
operator does not have a densely defined inverse. This seems to be a
severe obstacle, but it turns out that well-defined quantizations do
exist \cite{InvScale}, just as well-defined matter Hamiltonians exist
in the full theory \cite{QSDV}. In quantizations, the most obvious
procedure is not always the successful one, and also here one has to
start from alternative expressions for $a^{-3}$ which are identical
classically but lead to different quantizations.

One can rewrite $a^{-3}$ as, e.g.,
\begin{equation}
 a^{-3}= \left(\frac{3}{8\pi G{lj}({j}+1)(2{
 j}+1)}\sum_{I=1}^3{\rm tr}_{j}(\tau_I
 h_I\{h_I^{-1},|p|^{l}\})\right)^{3/(2-2{ l})}
\end{equation}
with parameters $0<l<1$ and $j\in\frac{1}{2}{\mathbb N}$, using only
positive powers of $p$ and ``holonomies'' $h_I=e^{c\tau_I}$ of the
connection component $c$. Inserting the basic operators and turning
the Poisson bracket into a commutator, this can directly be quantized
to a well-defined operator with eigenvalues \cite{Ambig,ICGC}
\begin{equation}
\widehat{d(a)}_{\mu}^{(j,l)} =
\left(\frac{9}{\ell_{\rm
P}^2lj(j+1)(2j+1)} \sum_{k=-j}^j
k|p_{\mu+2k}|^l\right)^{3/(2-2l)}\,.
\end{equation}
As one can see, the eigenvalues do depend on the parameters $j$ and
$l$, unlike the classical expression. Rewriting in the above manner
thus introduces ambiguities as it is expected for the quantization of
any non-basic operator. Important properties are, however, robust. For
instance, for any choice of the parameters we obtain the classical
behavior of $a^{-3}$ at large values $\mu\gg j\lP^2$, a peak around
$\mu_*=j\lP$ and decreasing behavior on small scales reaching exactly
zero for $\mu=0$.

For larger $j$, the sum in the eigenvalues contains many terms, and it
can be approximated by viewing it as a Riemann sum of an integral. In
this way, we obtain the {\em effective density}
\begin{equation} \label{deff}
 d(a)^{(j,l)}_{\rm eff}:=
 \widehat{d(a)}_{\mu(a^2)}^{(j,l)}= a^{-3} p_{{
 l}}(3a^2/ {j}\ell_{\rm P}^2)^{3/(2-2{ l})}
\end{equation}
with $\mu(p)=6p/\ell_{\rm P}^2$ and
\begin{eqnarray}
\textstyle p_l(q) &\textstyle =& \textstyle
\frac{3}{2l}q^{1-l}\left( \frac{1}{l+2}
\left((q+1)^{l+2}-|q-1|^{l+2}\right)\right.\nonumber\\
 && \textstyle\qquad- \left.\frac{1}{l+1}q
\left((q+1)^{l+1}-{\rm sgn}(q-1)|q-1|^{l+1}\right)\right) \,.
\end{eqnarray}
The approximation becomes better for larger $j$ as the number of terms
in the Riemann sum increases, which can also be seen in
Fig.~\ref{Dens}.

\begin{vchfigure}[htb]
 \includegraphics[width=.5\textwidth]{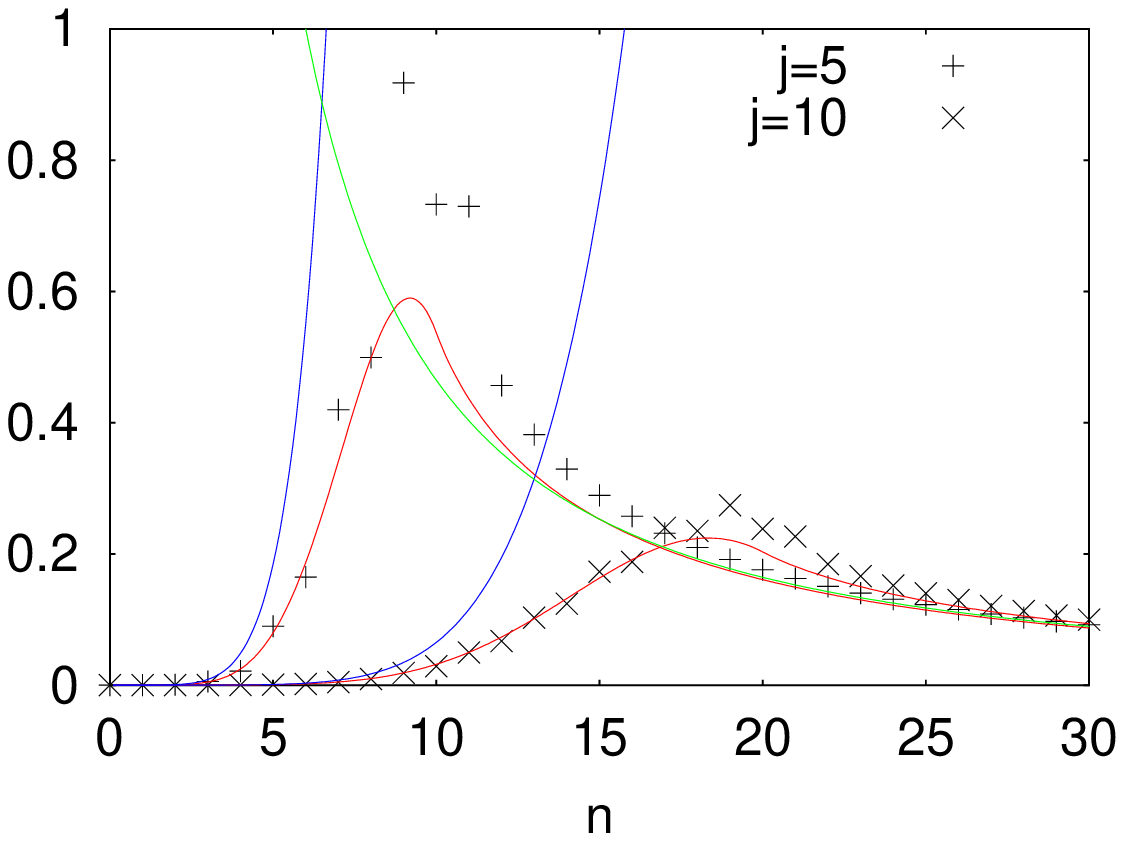}
\vchcaption{Discrete sets of eigenvalues of
  $\widehat{d(a)}_{\mu}^{(j,l)}$ for two values of $j$ and $l=3/4$
  compared to the classical behavior $a^{-3}$, the approximations by
  $d(a)_{\rm eff}^{(j,l)}$ and small-$a$ power-law approximations.}
\label{Dens}
\end{vchfigure}

Analogous constructions of $d(a)$ exist, e.g.\ $d(a_1,a_2,a_3)$ in
anisotropic models with diagonal metric components $g_{II}=a_I^2$. In
contrast to an isotropic context, these functions are not required to
be bounded for all configurations, i.e.\ all $a_I$ (see, e.g.,
\cite{DegFull}). The isotropic situation is very special in the way
the approach to the classical singularity at $a=0$ happens for which
there is only one possible trajectory in minisuperspace (which can
only be followed with different rates in coordinate time $t$). In less
symmetric models, there is much more freedom in the approach such as
$a_I(t)$ in anisotropic models, and not all possible configurations
are realized along the generic approach. Also in the full theory,
degenerate configurations exist on which inverse volume operators have
unbounded expectation values
\cite{BoundFull}.  At this point, dynamical information has to be used
to find out if curvature remains bounded along effective trajectories
of universe models.  

\begin{vchfigure}[htb]
  \includegraphics[width=.6\textwidth]{Pot4.eps} \vchcaption{Curvature
  potential on the minisuperspace of a diagonal Bianchi IX model with
  one metric component held fixed. Curvature is unbounded on
  minisuperspace, in particular at large anisotropies. Close to
  classical singularities (center and diagonal lines at $x=0$ or
  $y=0$), however, it remains bounded.}
\label{Pot4}
\end{vchfigure}

This can be done in homogeneous models such as the Bianchi IX model
\cite{Spin} where curvature is unbounded on all of minisuperspace
(Fig.~\ref{Pot4}) but bounded along the effective and even quantum
evolution as given by the difference equation.  On smaller scales,
also deviations from the classical approach happen which imply
non-chaotic behavior as the curvature walls start to break down from
quantum effects \cite{NonChaos,ChaosLQC}. From this, one can draw
conclusions for the generic inhomogeneous approach and structure
formation using the BKL picture \cite{BKL}, but for definitive
conclusions inhomogeneous models have to be studied in more detail and
in particular perturbative schemes for structure formation.

\subsection{Effective equations}

Difference equations, in particular partial ones in the presence of
matter fields or ansiotropies and inhomogeneities, are difficult to
deal with, and also interpretational issues of the wave function arise
at this level. While the full quantum setting is needed to discuss
singularities and the discrete evolution in their vicinity, on larger
scales it is more direct to use effective equations of classical
type. Those equations are differential equations in coordinate time,
but they are amended by modifications to capture quantum effects. This
is analogous to effective action techniques, and indeed there is a
general scheme for effective classical approximations of quantum
systems \cite{Perturb,Effective} which can be applied to quantum
cosmology in a canonical formulation and reproduces the usual
effective action results \cite{EffAc} obtained by expanding around
free field theories or the harmonic oscillator \cite{Effective}.

This scheme is based on a geometrical formulation of quantum mechanics
\cite{Schilling} where the Hilbert space is interpreted as an infinite
dimensional vector space with additional structure. The infinite
dimensionality, even for a mechanical system, is the crucial
difference between quantum and classical physics. There are thus
additional quantum degrees of freedom which in some regimes have to be
taken into account (often appearing as higher derivative terms in
effective actions). From the inner product of the Hilbert space one
obtains a symplectic structure as well as a metric on the vector space
such that it becomes K\"ahler. In addition, the quantum Hamiltonian
defines a flow on the K\"ahler manifold. With the symplectic structure
and the flow one has a canonical system equivalent to the quantum
system and in general involving all infinitely many degrees of
freedom. In some cases, however, it is possible to approximate the
flow by a dynamical system on a finite dimensional subspace of the
full space, giving rise to an effective classical system. This may
involve only the classical degrees of freedom, but with correction
terms in the evolution, or also additional ones related to, e.g., the
spread of wave functions.  The metric on the K\"ahler space is needed
only for the measurement process in quantum mechanics and issues such
as the collapse or overlap of wave functions. Since there is no
external observer in quantum cosmology and only one wave function, the
metric may be dropped which means that one could, as mentioned in
\cite{CK}, weaken the Hilbert space structure required
usually. Indeed, the geometrical formulation of quantum mechanics
provides a unified scheme for generalizations of quantum mechanics
\cite{Schilling}.

In loop quantum cosmology, several related methods have been applied
in order to derive effective terms for equations of motion, although a
complete derivation is still unfinished. The geometrical scheme is
developed for quantum cosmology in \cite{Perturb} with an
asymptotic series of correction terms for isotropic models. Leading
orders of such terms have also been derived in \cite{SemiClassEmerge}
and with WKB techniques in \cite{EffHam}. Some of these terms have
been used in applications
\cite{Time,GenericBounce,AmbigConstr,SemiClassEmerge}, but many
effects already show up at the level where additional degrees of
freedom are considered only to the lowest order. The order of
differential equations is then the same as classically, but terms in
the Hamiltonian do change. In particular, the quantum Hamiltonian
requires, in the presence of matter or other curvature terms, a
quantization of $a^{-3}$ which is modified at the quantum level.
Effectively, $a^{-3}$ in the matter Hamiltonian is replaced by
$d(a)$ from (\ref{deff}) \cite{Inflation}:
\begin{equation}
  H_{\phi}(a)=\frac{1}{2}d(a) p_{\phi}^2+a^3V(\phi)
\end{equation}
which behaves differently from the classical expression for
$a < \sqrt{j/3} \ell_{\rm P}$.

From the matter Hamiltonian we obtain Hamiltonian equations of motion
for the matter field, which now change on small scales \cite{Closed}.
For a scalar, we have the {\em effective Klein--Gordon equation}
\begin{equation} \label{EffKG}
\ddot{\phi}=\dot{\phi}\,\dot{a}\frac{\mathrm{d}\log {
    d(a)}}{\mathrm{d} a}-a^3{d(a)}V'(\phi)
\end{equation}
where the usual friction term $-3\dot{\phi}\dot{a}/a$ changes its
form. Back-reaction from matter on geometry is encoded in the
Friedmann equation which also changes. Substituting the effective
matter Hamiltonian, we obtain the
     {\em effective Friedmann equation}
\begin{equation} \label{EffFried}
 a(\dot{a}^2+k^2)={\textstyle\frac{8\pi}{3}}G
\left({\textstyle\frac{1}{2}}{d(a)}\, p_{\phi}^2+a^3
V(\phi)\right)
\end{equation}
and from the equations of motion generated by the constraint the
{\em effective Raychaudhuri equation}
\begin{equation} \label{EffRay}
  \frac{\ddot{a}}{a}= -\frac{8\pi G}{3}\left( a^{-3}{d(a)}p_{\phi}^2 
\left(1-{\textstyle\frac{1}{4}}a\frac{\md
\log(a^3{ d(a)})}{\md a}\right) -V(\phi)\right)\,.
\end{equation}
In the latter, the modification leads to an entirely new term.

\section{Phenomenology}
  
These equations, all resulting from a single modification in the
matter Hamiltonian implied by effects of the basic quantum
representation, are the starting point for phenomenology of loop
cosmology based on effective equations. Additional corrections,
related to quantum fluctuations and additional quantum degrees of
freedom, have not yet been studied systematically but are expected to
be important only on very small scales. Modifications in the matter
Hamiltonian, on the other hand, are non-perturbative and can be
shifted into regimes of larger scales just by choosing a large value
for $j$. This allows one to study its implications in isolation from
other correction terms, even though one does not expect $j$ to be very
large. Additional corrections \cite{EffHam,SemiClassEmerge} and
similar parameters \cite{AmbigConstr} also arise for the gravitational
part of the constraint and become relevant when the matter density is
large (of Planck size).

The main modification can then easily be interpreted intuitively by
viewing the Friedmann equation as the energy equation of a classical
mechanics system with a potential determined by the matter
Hamiltonian. Classically, the matter Hamiltonian is usually decreasing
as a function of $a$ as a consequence of the kinetic term. This is a
consequence of the fact that classical gravity is always attractive.
When $a^{-3}$ is replaced by $d(a)$ as in Fig.~\ref{Dens}, however,
the slope of the potential is flipped on small scales and it becomes
increasing. Thus, the direction of the force changes and (quantum)
gravity on small scales receives a {\em repulsive} contribution. With
this picture, one can easily imagine that characteristic effects can
arise such as the prevention of collapse into a singularity by a
bounce or the acceleration of expanding evolution to an inflationary
era.

\subsection{Bounces}

For a bounce to be realized, we need to find a time where $\dot{a}=0$
and $\ddot{a}>0$. The first condition can be checked with the
effective Friedmann equation (\ref{EffFried}) where, for $\dot{a}=0$
to be possible, the positive kinetic term must be compensated by
either a positive curvature term $k=1$
\cite{BounceClosed,BounceQualitative} or a negative scalar potential
$V(\phi)<0$ \cite{Cyclic}. The second condition at solutions for
$\dot{a}=0$ is then controlled by the effective Raychaudhuri equation
(\ref{EffRay}) which for classical solutions gives only negative
$\ddot{a}$, i.e.\ recollapse points.

\begin{figure}[htb]
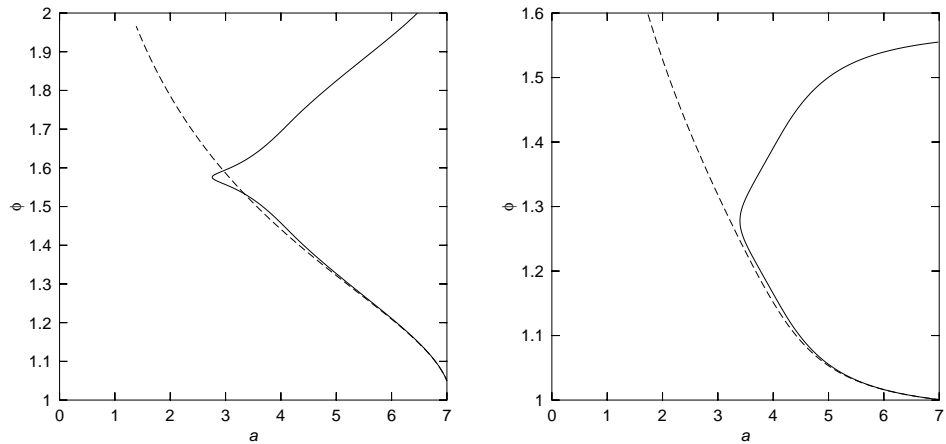

  \includegraphics[width=.4\textwidth]{bouncecurv.eps}\hspace{5mm}
 \includegraphics[width=.4\textwidth]{bounceneg.eps}
\caption{Numerical solutions (solid lines) for bounces in models with
  positive spatial curvature (left) and a negative potential (right),
  respectively, compared to the classically singular solutions (dashed).}
\label{bounces}
\end{figure}

When a turning point falls in the modified regime, on the other hand,
the additional term in the effective Raychaudhuri equation as well as
modified matter behavior through the effective Klein--Gordon equation
(\ref{EffKG}) can change the picture and imply bounces. This is not
realized in all cases, but happens generically and without the need
for special potentials in contrast to the classical
situation. Numerical solutions for two examples are given in
Fig.~\ref{bounces}. These effective bounces can be seen as a
consequence of a repulsive gravitational force, or equivalently of
negative pressure
\begin{equation}
 P=-\frac{\partial H}{\partial V}=-\frac{1}{3a^2}\frac{\partial
   H}{\partial a}<0
\end{equation}
which changes sign on scales where the matter Hamiltonian is increasing.

The modification in the matter Hamiltonian alone does not give rise to
bounces in all cases, such as a flat model with positive potential. A
universe then still collapses to small sizes closer and closer to the
deep quantum regime. Eventually, its dynamics has to be described by
the basic difference equation which is non-singular, but at small
sizes also additional corrections become important in effective
equations. Those corrections are indeed directly related to the
underlying discreteness or quantum fluctuations of geometry, and
appear on very small scales. This, then, provides bounces more
generically \cite{GenericBounce,KasnerBounce,Time,APS}.

\subsection{Inflation}

Negative pressure is also required for inflation where $\ddot{a}>0$
can now follow without special conditions for the potential or the
initial values of an inflaton field. In fact, inflation in the general
sense of accelerated expansion now happens generically and even
without any potential at all \cite{Inflation} as illustrated in
Fig.~\ref{Infl}. It is enough that the kinetic term in the effective
Friedmann equation becomes increasing as a function of $a$ which is
always realized on small scales.

\begin{vchfigure}[htb]
  \includegraphics[width=.6\textwidth]{InflAeff.eps}
\vchcaption{Numerical solution with inflationary behavior (right) in
  the regime of modified effective densities (left). Dashed lines
  indicate the range of sizable modifications.}
\label{Infl}
\end{vchfigure}

The precise manner depends on the ambiguity parameter $l$ for which we
obtain $d(a)\sim {a^{3/(1-l)}}$ on small scales which is always
increasing due to $0<l<1$. This form determines, for a vanishing
potential, the type of inflation which is super-inflationary (with
equation of state parameter $w=-1/(1-l)<-1$). For more realistic
models with a potential, however, the behavior is driven very close to
exponential inflation thanks to the $a$-dependence of the potential
term \cite{GenericInfl}.  From the behavior of matter in such an
inflating background one can then derive potentially observable
effects.

We first look at the kinetic term driven inflation which happens for
any matter field and thus can eliminate the need for an inflaton.
During accelerated expansion in this regime, structure can be
generated with less fine tuning than in inflaton models. As
preliminary calculations indicate \cite{PowerLoop}, the resulting
spectrum is nearly scale invariant \cite{GenericInfl}, and quantum
effects can also be used to give arguments, based on
\cite{InflCutOff,InflCutOff2}, for a small amplitude in agreement with
observations.  Details of the spectrum such as the running of the
spectral index do depend on ambiguity parameters such that they may be
restricted by forthcoming detailed observations. In contrast to single
field inflaton models, the spectral index would be a little larger
than one as a consequence of super-inflation, i.e.\ the spectrum is
slightly blue, which can be a characteristic signature.

If this phase alone would have to be responsible for the generation of
all structure, an extremely large value for the parameter $j$ would be
required. Moreover, the spectrum would then be blue on all scales
which is ruled out by observations. As we will see now, however, the
first inflationary phase will always be followed by slow-roll phases
because matter (or inflaton) fields are driven away from their
potential minima while the modified Klein--Gordon equation is
active. Subsequent phases can thus serve to make the universe large
enough and provide a sufficient amount of $e$-foldings, while visible
structure can have been generated in the first, quantum
phase. Attractive features of this class of scenarios are that
potentials for matter fields driving late-time slow-roll phases do not
need to be special and that visible structure would have been
generated by the quantum phase, thus giving potentially observable
quantum signatures.

\subsection{Scalar dynamics}

\begin{figure}[htb]
  \includegraphics[width=0.95\textwidth]{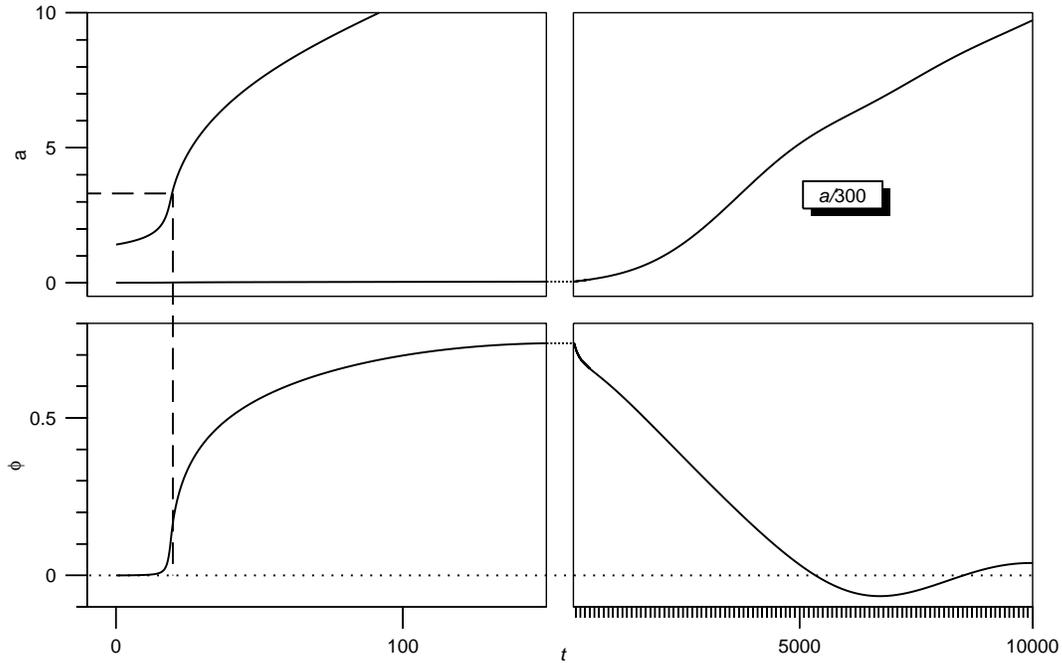}
\caption{Initial push of a scalar field $\phi$ up its potential with
  subsequent slow-roll to and oscillations around the potential
  minimum at $\phi=0$.}
\label{Push}
\end{figure}

In the effective Klein--Gordon equation (\ref{EffKG}) the classical
friction term of an expanding universe, used for slow-roll behavior,
changes sign and turns into an {\em antifriction} term on small scales
\cite{Inflation}. Matter fields in this regime are then excited and
can move up their potential walls even if they start close to minima
at small momenta. After antifriction subsides, they will continue to
roll up the walls but be slowed down by the then active friction.
Eventually, they turn around and roll down the potential slowly. In
this manner, additional phases of inflation are generated. The whole
history is shown in Fig.~\ref{Push} with a rapid initial phase
containing the push of $\phi$ up its potential and the loop
inflationary phase together with the later slow-roll stage. A solution
for the scale factor showing both inflationary phases in the same plot
is given in Fig.~\ref{SlowRolla}.

If $\phi$ is an inflaton such as that of chaotic inflation with a
quadratic potential, initial conditions are provided for a long phase
of slow-roll inflation. But the slow-roll conditions are not satisfied
in all stages, in particular not around the turning point of the
inflaton. Here, $\dot{\phi}$ vanishes or is very small such that the
slow-roll condition $\ddot{\phi}\ll H\dot{\phi}$ cannot be satisfied.
If the second phase is responsible for structure formation, structure
on large scales, generated in early stages of the slow-roll phase,
will differ from usual scenarios. This can in particular contribute to
a suppression of power on large scales \cite{InflationWMAP}. If the
last inflationary phase takes too long these effects are not visible,
but one can estimate the duration from the inflaton initial values one
typically gets through the antifriction mechanism. As it turns out,
one often obtains observable effects, i.e.\ there is a sufficient
amount of inflation but not too much for washing away quantum
gravity effects \cite{Robust}. This aspect is even enhanced if one
looks at the evolution at the level of difference equations
\cite{ASV}.

\begin{vchfigure}[htb]
  \includegraphics[width=.5\textwidth]{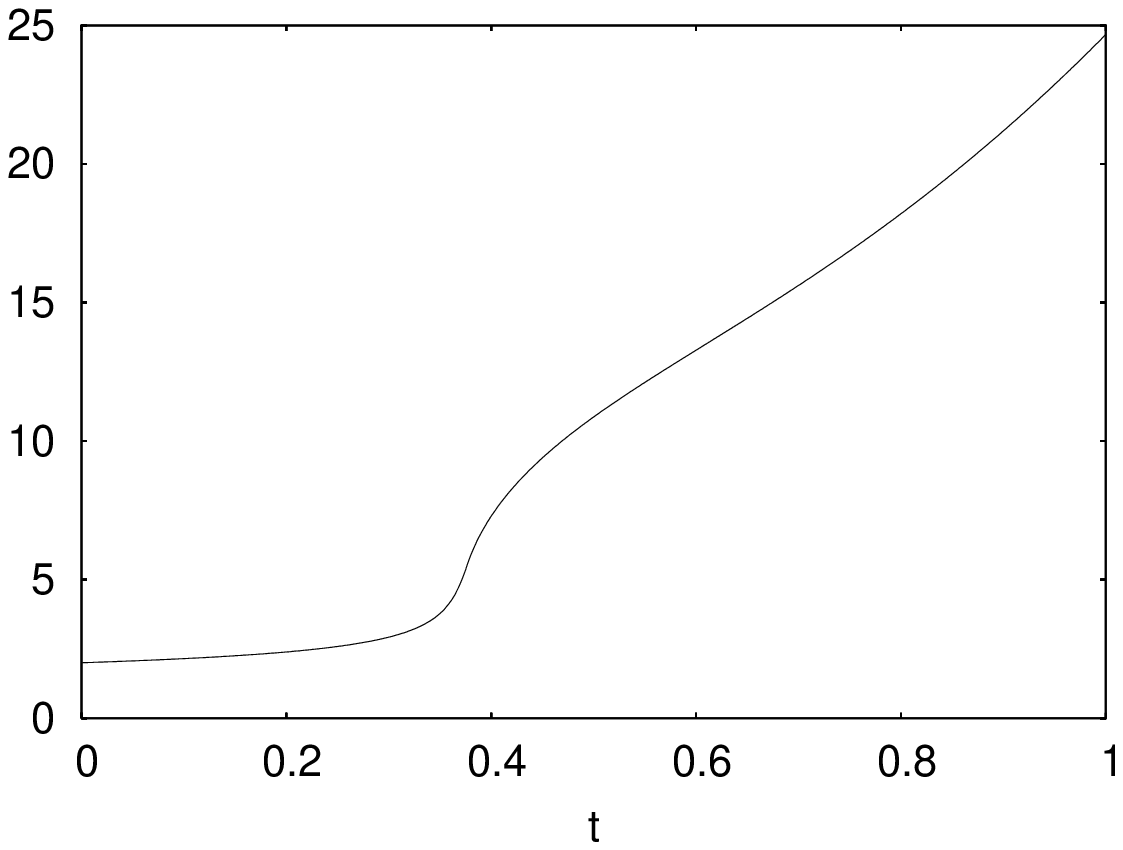}
\vchcaption{Numerical solution showing the first inflationary phase of
  loop cosmology and the beginning of the ensuing slow-roll phase
  after the scalar is pushed up its potential.}
\label{SlowRolla}
\end{vchfigure}

\subsection{Combinations of different phases}

So far we have considered individual bounces or inflationary phases.
Depending on details of the matter system such as potentials, many
combinations are possible and often of interest for model building.
If one combines several bounces in models with a classical recollapse,
{\em oscillatory models} result \cite{Oscill}. The behavior can then
gradually change from cycle to cycle and eventually, after small
changes have added up, result in a qualitative change in the universe
behavior.

This is realized, for instance, in a new version of the emergent
universe which was originally devised as a non-singular inflationary
model with positive spatial curvature \cite{Emergent}. The
singularity is avoided by starting the model close to a static
Einstein space in the infinite past which, with a suitable potential,
develops into an inflationary phase. The initial state is, however,
very special because the static Einstein space is unstable. This
changes if effective equations of loop cosmology are used: there are
new static solutions on small scales which, in contrast to the
classical ones, are stable \cite{EmergentLoop}. Starting close to
those solutions will lead to a series of cycles of a small universe
which can gradually change due to the motion of matter fields in their
potential.

In this manner, one obtains a whole history for a universe evolving
through different phases of contraction and expansion in a cyclic but
not necessarily periodic manner. However, the cycles are usually short
and do not automatically give rise to a large universe. On the other
hand, matter fields move in their potential during the evolution, and
close to bounce points antifriction effects can help to bring matter
fields far up their potentials or over potential barriers. If a
suitable part of the potential is encountered during this process, the
conditions can be right for the start of a slow-roll regime of
sufficient duration for a large universe to emerge \cite{InflOsc}. In
fact, this can be obtained in the emergent universe scenario from a
simple initial state evolving, after many cycles, to an inflationary
phase for the structure formation we need for our universe
\cite{EmergentLoop,EmergentNat}. A numerical solution for this
scenario is illustrated in Fig.~\ref{Emergent}. The characteristic
feature is that it is based in an essential manner on
closed spatial slices of positive curvature which, if the inflationary
phase is not too long, may be detectable in the near future
\cite{EmergentLoop}.

\begin{vchfigure}[htb]
  \includegraphics[width=.5\textwidth]{EmergentWave.eps}
  \vchcaption{Stroboscopic density plot of solutions in the classical
  phase space (arbitrary units) with initial cyclic behavior and an
  eventual inflationary phase. To illustrate the time behavior,
  Gaussians have been added peaked at discrete coordinate time
  intervals.}
\label{Emergent}
\end{vchfigure}

When different matter sources are present, such as different fields or
a fluid in addition to a scalar field, other possibilities arise.
There can still be fixed points which allow cyclic behavior around
them, but the position can now depend on the field values. In
particular, cyclic motion of the fixed points themselves is possible
which implies that the evolution is double-cyclic with small cycles
around a fixed point superposed to the cyclic motion of the fixed
point \cite{LoopFluid}. There are several new possibilities which are
being investigated in dynamical system approaches. In this context it
is in particular of interest when a {\em graceful entrance} into an
inflationary phase can arise in the presence of different matter
sources, i.e.\ how generically initial conditions become right to
start a sufficiently long slow-roll phase.

Cyclic behavior is also studied often in the context of brane
collisions. While loop quantum gravity and cosmology are difficult to
formulate in this higher-dimensional setting,\footnote{There is,
however, an interesting duality to brane-world models if conditions
for the shape of $d(a)$ are relaxed \cite{LoopBraneDual}.} one can
simply take potentials motivated from brane scenarios \cite{Roy} and
study implications with loop modifications.  The interpretation of the
scalar is then as a radion field, i.e.\ the distance between two
branes. A common characteristic property is the possibility of
negative potentials which, as we have seen earlier, allow bounces in
loop cosmology.

\begin{figure}[htb]
  \includegraphics[width=.47\textwidth]{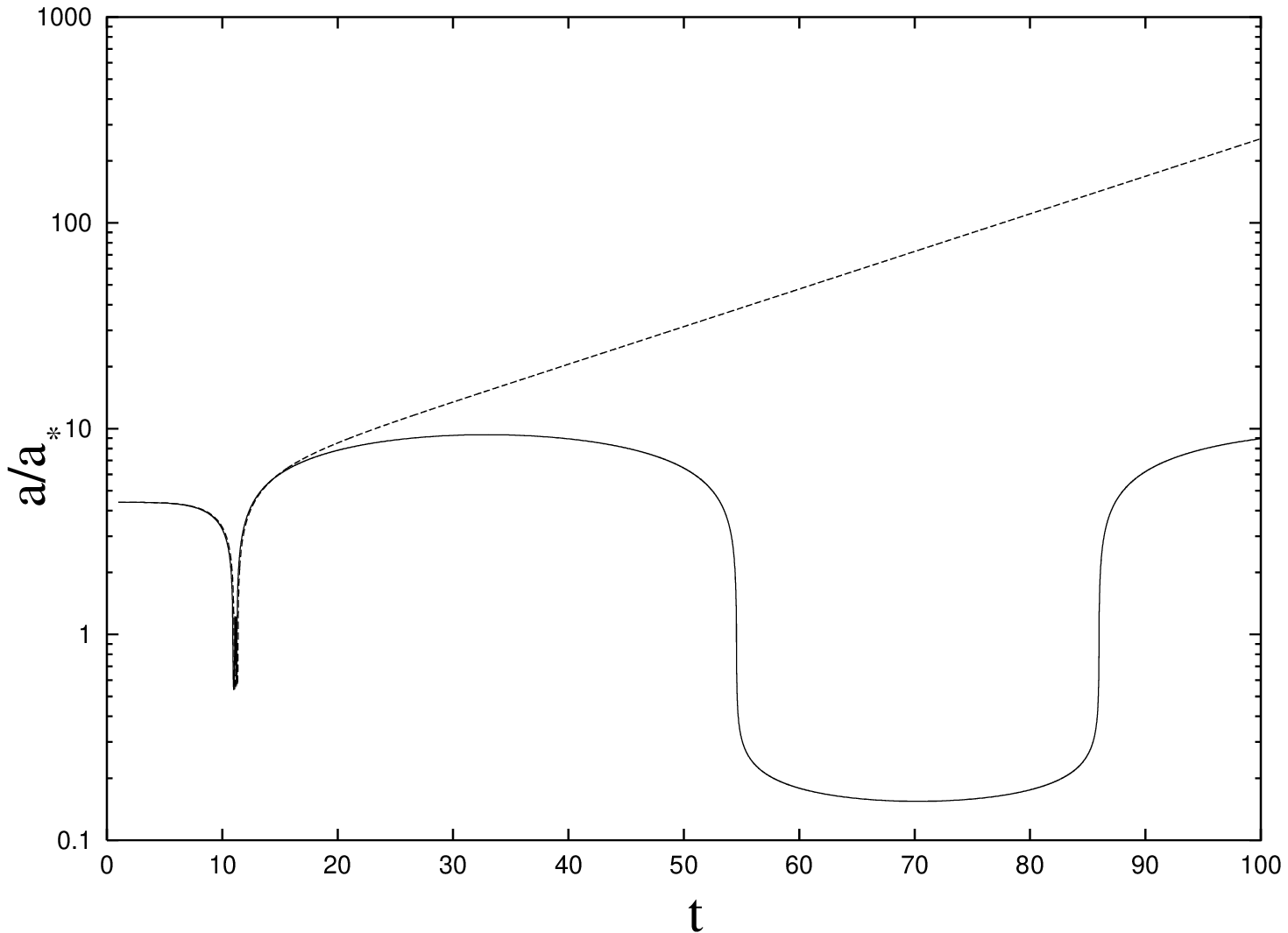}
 \includegraphics[width=.5\textwidth]{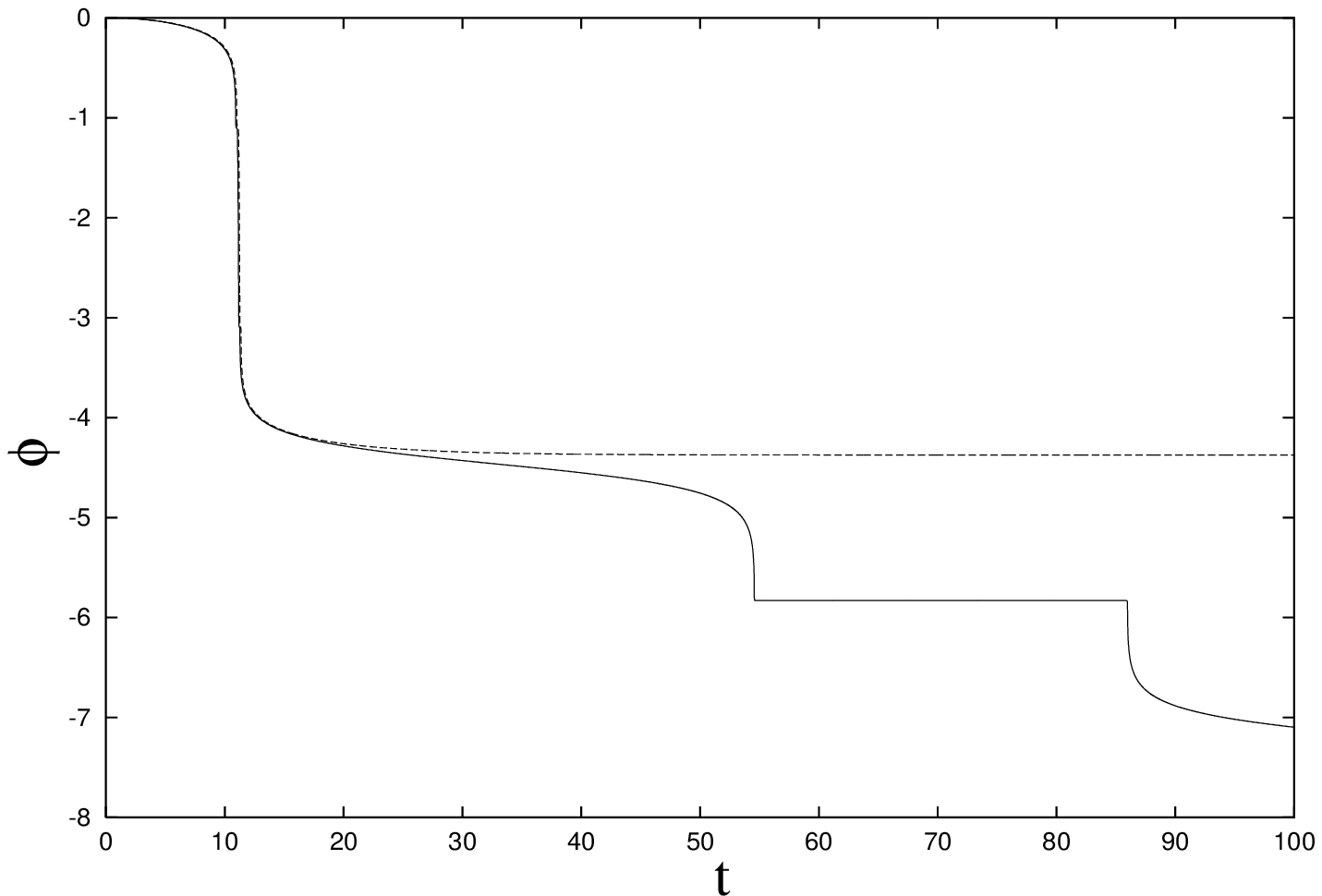}
\caption{Bouncing numerical solution for the scale factor (left) in a brane 
  potential with a monotonic decrease of the scalar (right). During
  bounces, the scalar is only slowed down due to antifriction in the
  effective Klein--Gordon equation, but it cannot turn around. A
  turnaround can be realized only if the scalar encounters positive
  values of its potential, which generically leads to inflation
  (dashed).}
\label{BraneSol}
\end{figure}

Loop cosmology then provides a non-singular bounce for such a brane
model, and one can check how easy it is to realize properties assumed
in other models where a mechanism for singularity removal was not
known. In some of those scenarios, for instance, it was assumed that
the scale factor bounces and simultaneously the scalar field turns
around \cite{Ekpyrotic}. Only then does one really have a
model where the branes first approach each other and then bounce
off. This turns out to be impossible in loop cosmology based on
(\ref{EffFried}) where bounces are realized with a negative potential
but, as illustrated in Fig.~\ref{BraneSol}, the scalar cannot change
direction \cite{Cyclic}.  The only possibility for $\phi$ to turn
around is if it encounters a region of positive potential, but then
one automatically obtains an inflationary phase and the scenario is
not different from those described before. Similar results can be
obtained when bouncing solutions realized with higher curvature
corrections are used
\cite{BounceString}.

\section{Conclusions}

Through its background independent and non-perturbative quantization
scheme, loop quantum gravity and cosmology are well-equipped for
extreme physical situations as they are realized close to classical
singularities. With symmetric models, those situations can now be
dealt with in many cases including cosmology and black holes. Many
examples exist for non-singular models by the same general mechanism,
and also conceptual problems can be solved. In addition, many
phenomenological applications arise from a few basic effects.

These effects have, in all cases, been known first from mathematical
considerations and then transferred to and evaluated explicitly in
models. The discussed effects then resulted automatically, rather than
being looked for with particular applications in mind. Moreover, a few
effects going back to properties of the basic representation dictated
by background independence suffice to cover a plethora of physical
applications in different areas.

This fact is encouraging for the viability of the whole framework, but
it is still important to understand the relation between models and
the full theory as completely as possible. Some properties can
directly be related, and others have to be derived after analogous
constructions. There are no known contradictions, but at the dynamical
level also no proof of, e.g., singularity removal without assuming
symmetries. It is, however, clear already that the presence of local
physical degrees of freedom is by itself no obstacle to non-singular
quantum evolution.

Also models, in particular inhomogeneous ones, have to be developed in
more detail. At the fundamental level one then has to deal with many
coupled partial difference equations, for which even the formulation
of a well-posed initial or boundary value problem can be
difficult. For solutions one will have to refer to new numerical
techniques which are being developed in investigations of numerical
quantum gravity. Effective equations would also help considerably in
understanding inhomogeneous situations, but their derivation is much
more complicated than in homogeneous models and so far not completed.

At the same time, properties of the full theory are being understood
better. There is thus an approach from two sides, by weakening
symmetries in models to get closer to the full setting and by starting
to understand full configurations which can be argued to be close to
states considered in a symmetric model. A third direction will, in the
future, be provided by observations and their relation to
phenomenological results. In this way, confidence in physical
effects derived with loop methods will be strengthened and can
eventually be compared with observations.

\end{document}